\documentclass[a4paper]{llncs}
\usepackage{llncsdoc}
\usepackage{url}

\begin{document}
\mainmatter
\title{Creating Textual Language Dialects Using Aspect-like Techniques}
\author{Andrey Breslav\thanks{Main results presented in this paper are accepted for publication in journal Vestink SPbSU ITMO. The work is partially supported by Saint-Petersburg State grant \#3.11/4-05/55}}
\institute{St. Petersburg State University of IT, Mechanics and Optics\\
\email{abreslav@gmail.com}
}
\authorrunning{ }
\maketitle

We present a work aimed on efficiently creating textual language dialects and supporting tools for them (e.g. compiler front-ends, IDE support, pretty-printers, etc.). A dialect is a language which may be described with a (relatively small) set of changes to some other language. For example we can consider SQL dialects used in DB-management systems. 

The need in creating dialects is witnessed by existence of extensible languages like LISP and many more \cite{PLOT,Nemerle,REBOL,Converge}. 
There also are special tools supporting dialect development for C \cite{xtc} and Java \cite{Polyglot}.

These approaches allow extending languages efficiently, but it took a lot of work to make it possible for each single language. The problem we are addressing here is {\it how can we do it for arbitrary languages with not so much effort?} 

If we need only a syntax checker for a language, creating dialects for it is easy. We have a grammar describing the language syntax and all we need is to modify this grammar. This can be done easily by a transformation which may be formulated in terms of aspect application (analogous to AOP \cite{AOP}), we call such transformations {\it syntactical aspects}. By pattern matching aspects identify places in the source grammar that we need to modify and provide instructions for such modifications: insert something before or after the specified point, or replace the object with some other one.
For example, assume we have a grammar for arithmetical expressions for real numbers:
\begin{verbatim}
sum : mult ('+' mult)*;
mult : factor ('*' factor)*;
factor 
    : REAL
    : '(' sum ')'
    ;
\end{verbatim}
We may want to replace reals with integers and introduce variables in this grammar, this could be done by modifying rules for {\tt factor} as follows:
\begin{verbatim}
factor 
  $production=|: REAL
    @REAL.instead = << INT >>
    @production.after = <<: ID >>
  ;
\end{verbatim}
This replaces one token reference with another and inserts a new production rule. Many real-world grammar transformations which are needed for dialect creation are easily expressible in this manner.

Things become less obvious when we need something more than a parser. Assume we are working with a language front-end that supports parsing error recovery, semantic checking and abstract syntax tree creation. Now we need to transform the whole system, not only the grammar, and changes to all parts must be coherent. To achieve this we can observe that in most cases a structure of front-end features is highly dependent on the grammar structure -- high-level descriptions of features like semantic analysis or error-recovery are structured the same way as the grammar is. If we could reuse the grammar structure in these definitions, namely, provide explicit connection of definition parts to grammar objects, we would be able to transform the whole system coherently since all the structural changes would be induced by grammar changes.

We propose an approach which is based on this idea. In this approach all the features of the system are described by high-level models, which are attached to corresponding grammar objects. The working code is generated using information from these models. Technically model data is divided into name-value pairs (where values may be of different types) and attached to grammar objects with aspect-like rules. We call this {\it metadata aspects}. The way of assigning metadata enables high modularity since we do not need to specify all metadata in a single definition, but we are able to define a separate definition for each aspect of the system.

After metadata is assigned, all the changes to grammar structure automatically rearrange it, which is ensured by special integrity checkers. This makes dialect development very easy since in many cases all we need to create a dialect is to write a syntactical aspect to transform a grammar.

\bibliographystyle{unsrt}

\end{document}